\documentclass[twocolumn,english,prl,aps]{revtex4-1}
\usepackage[T1]{fontenc}
\usepackage[latin9]{inputenc}
\usepackage{units}
\usepackage{amsmath}
\usepackage{amssymb}
\usepackage{graphicx}
\usepackage{esint}

\makeatletter

\providecommand{\tabularnewline}{\\}

\usepackage[caption=false]{subfig}
\usepackage{babel}

\@ifundefined{showcaptionsetup}{}{%
 \PassOptionsToPackage{caption=false}{subfig}}
\usepackage{subfig}
\makeatother

\usepackage{babel}
\begin{document}

\title{Time-reversal asymmetry without local moments via directional scalar
spin chirality}

\author{Pavan R. Hosur}

\affiliation{Department of Physics, Stanford University, Stanford, CA 94305, USA}
\begin{abstract}
Invariably, time-reversal symmetry (TRS) violation in a state of matter
is identified with static magnetism in it. Here, a directional scalar
spin chiral order (DSSCO) phase is introduced that disobeys this basic
principle: it breaks TRS but has no density of static moments. It
can be obtained by melting the spin moments in a magnetically ordered
phase but retaining residual broken TRS. Orbital moments are then
precluded by the spatial symmetries of the spin rotation symmetric
state. It can exist in one, two and three dimensions under different
conditions of temperature and disorder. Recently, polar Kerr effect
experiments in the mysterious pseudogap phase of the underdoped cuprates
hinted at a strange form of broken TRS below a temperature $T_{K}$,
that exhibits a hysteretic ``memory effect'' above $T_{K}$ and
begs reconciliation with nuclear magnetic resonance (which sees no
moments), X-ray diffraction (which finds charge ordering tendencies)
and the Nernst effect (which detects nematicity). Remarkably, the
DSSCO provides a phenomenological route for reconciling all these
observations, and it is conceivable that it onsets at the pseudogap
temperature $\sim T^{*}$. A testable prediction of the existence
of the DSSCO in the cuprates is a Kerr signal above $T_{K}$ triggered
and trainable by a current driven along one of the in-plane axes,
but not by a current along the other.
\end{abstract}
\maketitle

\section{introduction}

A quantum phase of matter that spontaneously breaks time-reversal
symmetry (TRS) invariably develops a finite density of moments. In
other words, there exists a set of total angular momentum operators
$\left\{ J_{i}\right\} $ such that $\left\langle G\left|\sum_{i}J_{i}\right|G\right\rangle $
is extensive in its ground state $|G\rangle$. Common examples contain
local spin moments, such as ferromagnets, spin density waves and other
spin textures. More complex ones include orbital moments, such as
loop current phases \cite{Varma1997,He2012}, anomalous Hall states
\cite{Haldane1988,Nagaosa2010}, and various chiral topological phases
\cite{Kalmeyer1987,Yao2007,Wen1989,Dhar2013,Zalatel2014,Kallin2009,Leggett1975,Osheroff1972,Osheroff1972a,Tsuruta2015}.
A property common to all these phases is that TRS is restored as soon
as the moments melt. Thus, the phrases ``spontaneous violation of
TRS'' and the ``formation of local moments'' are often used interchangeably.
Strictly speaking, though, this synonymy is incorrect because local
moments also disobey spatial symmetries. A natural question that follows
is, can we find a phase of matter that violates TRS but has no moments?
Such a phase could be pertinent to a long-standing problem in condensed
matter physics -- the pseudogap phase of the cuprate high temperature
superconductors -- which exhibits a Kerr effect \cite{Xia2008,Spielman1992,Spielman1990,Karpetyan2014},
indicating broken TRS, but shows no signs of magnetism in nuclear
magnetic resonance (NMR) experiments \cite{Wu2015}.

In this work, precisely such a phase of matter is introduced, called
the \emph{directional scalar spin chiral spin order }(DSSCO). The
DSSCO can be thought of as a state in which classical magnetic order
has melted due to quantum, thermal or disorder-driven fluctuations
-- so spin moments vanish -- but TRS-breaking has survived. Moreover,
orbital moments involving itinerant particles, if any, are forbidden
by the symmetries of the DSSCO. It is captured by an order parameter
of the form $\chi\sim\left\langle \boldsymbol{S}_{1}\cdot\boldsymbol{S}_{2}\times\boldsymbol{S}_{3}\right\rangle $,
where $\boldsymbol{S}_{1}$, $\boldsymbol{S}_{2}$ and $\boldsymbol{S}_{3}$
are spins on three sites in a straight line. Thus, it is reminiscent
of some other phases that involve spin chirality, such as those studied
in Refs \cite{Grohol2005,Lee2013}. The key difference is that the
chirally correlated spins there lie on the vertices of a triangle.
Hence, they break enough symmetries to permit a moment perpendicular
to its face, even if the moment on each site vanishes. In contrast,
the corresponding sites in a DSSCO are collinear, so no such current
is possible. The precise conditions in which the DSSCO can form depends
sensitively on the dimensionality of space. In particular, it exists
in one dimension (1D) at zero temperature ($T=0$) in clean systems,
in 2D at $T\neq0$ in clean systems, and in 3D at both $T=0$ and
$T\neq0$ only in the presence of weak random field disorder. The
3D DSSCO respects spin rotation symmetry (SRS) is respected only on
averaging over disorder configurations, and is the one most relevant
to the cuprates. Nonetheless, the term DSSCO will be used to denote
all the phases based on chiral spin ordering along a preferred direction
that break TRS but lack a density of moments.

One of the most enigmatic phases known in condensed matter is the
pseudogap phase of the underdoped cuprates. Recently, Kerr effect
measurements in this phase showed a signal below a certain temperature
dubbed $T_{K}$ \cite{Xia2008,He2011,Karpetyan2012,Karpetyan2014},
strongly suggesting that TRS is broken below it \cite{Halperin1992,Fried2014}.
However, the symmetries of the phase are distinct from that of an
ordinary magnet. Moreover, NMR Knight shifts -- usually an excellent
probe of magnetic order -- have not found any magnetic moments till
date \cite{Wu2015}. We will see that the Kerr effect in the DSSCO
in a clever experimental setup has precisely the same symmetries as
that in the cuprates, while the Knight shifts vanish identically.
Remarkably, a different scenario in a traditional setup permits a
route for reconciling several baffling behaviors experimentally observed
in the cuprates: (i) $C_{4}$ symmetry breaking above $T_{K}$ \cite{Daou2010}
but below the pseudogap temperature $T^{*}$, (ii) coincident onsets
of the Kerr effect and charge ordering tendencies \cite{Blackburn2013,Comin2015,LeTacon2014},
(iii) magnetic moments possibly undetectable by NMR \cite{Mounce2013},
and (iv) a hysteretic ``memory effect'' on heating beyond $T_{K}$
\cite{Xia2008}. It is unclear, however, if the microscopics of the
DSSCO -- especially the presence of random fields, which will be shown
to play a vital role shortly -- can apply to the cuprates. Moreover,
it does not explain the magnetism predicted by neutron scattering
\cite{Baledent2010,Bourges2011,Fauque2006,Li2008,Mook2008}. Nonetheless,
the phenomenology is rather appealing as it can capture many different
experiments in the cuprates.

\section{Directional scalar spin chiral order}

\begin{figure}
\subfloat[\label{fig:DSSCO-1D}]{\begin{raggedright}
\includegraphics[clip,width=0.45\columnwidth]{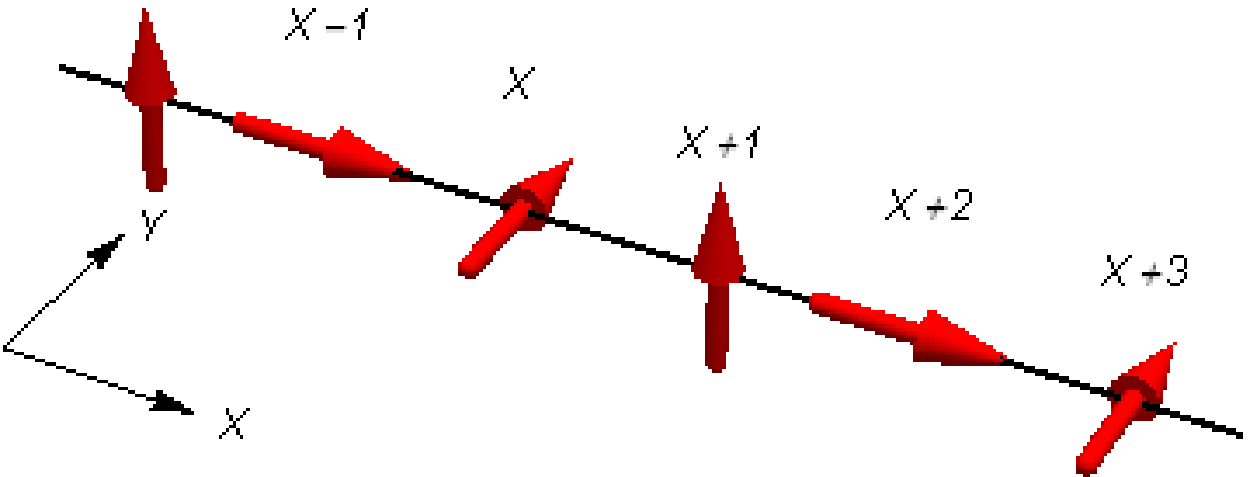} 
\par\end{raggedright}

}\subfloat[\label{fig:DSSCO-2D}]{\begin{raggedright}
\includegraphics[clip,width=0.45\columnwidth]{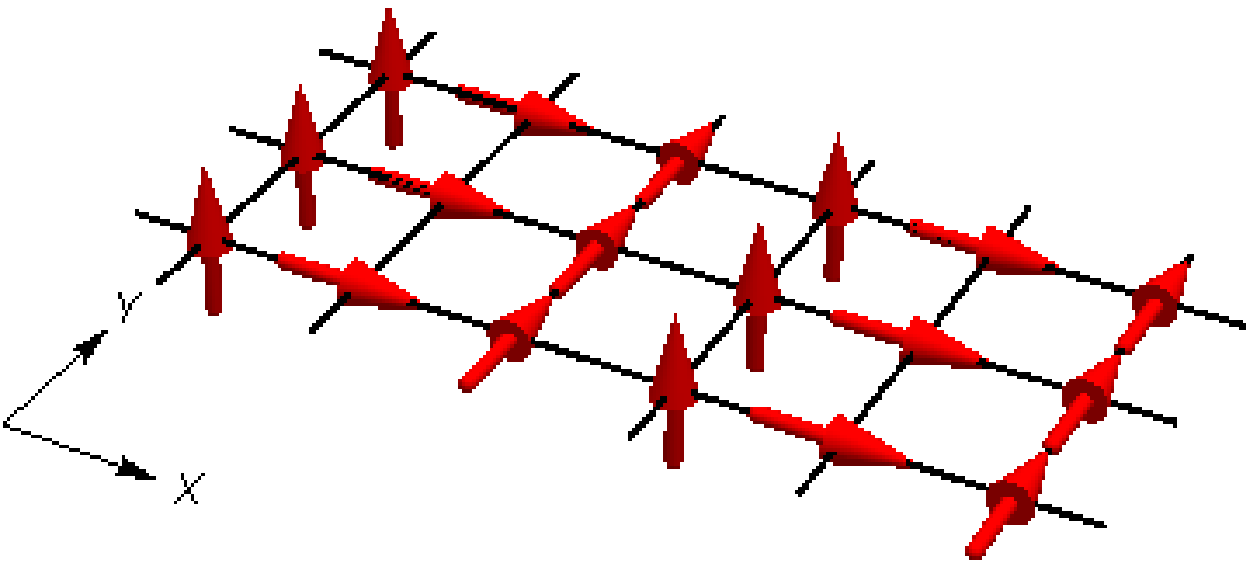} 
\par\end{raggedright}

}

\caption{Classical magnetic orders which form the DSSCO upon melting in 1D
(a) and 2D (b). Stacking identical 2D layers gives the precursor to
the 3D version of the DSSCO.\label{fig:magnetic-order}}
\end{figure}

Let us first sketch the 1D version of the DSSCO. Consider an ordering
of classical (large $S$) spins along a chain as shown in Fig. \ref{fig:DSSCO-1D}.
Here, spins on successive sites are frozen in the pattern $S_{x}S_{y}S_{z}S_{x}S_{y}S_{z}\dots$.
Such a pattern of magnetic moments obviously violates TRS and SRS;
in addition, it also breaks all reflection symmetries. A potential
order parameter for it is the pseudoscalar 
\begin{equation}
\chi=\frac{1}{L}\sum_{x}\left\langle \boldsymbol{S}(x-1)\cdot\boldsymbol{S}(x)\times\boldsymbol{S}(x+1)\right\rangle \label{eq:chi-1D}
\end{equation}
where $L$ is the chain length. Clearly, $\chi$ is an Ising order
parameter that distinguishes between right-handed ($S_{x}S_{y}S_{z}S_{x}S_{y}S_{z}\dots$)
and left-handed ($S_{z}S_{y}S_{x}S_{z}S_{y}S_{x}\dots$) sequences
of spins. These sequences transform into each other under time-reversal
($\boldsymbol{S}\to-\boldsymbol{S}$) or inversion ($x\to-x$). However,
$\chi$ is invariant under reflection about any plane containing the
chain as well as under a global rotation of all the spins, so it does
not fully capture the classical order. Let us assume that Fig. \ref{fig:DSSCO-1D}
depicts the ground state of a classical, local spin Hamiltonian that
preserves TRS and SRS and has no disorder. If the spins were quantum
objects instead, fluctuations would immediately restore SRS according
to the Mermin-Wagner theorem \cite{Mermin1966}. In contrast, TRS
and reflection symmetry are discrete and can hence, remain broken.
A closer inspection reveals that the resultant state disrespects TRS
and inversion symmetry, but is invariant under translation and spin
rotation. Therefore, it is faithfully captured by the order parameter
$\chi$. This state is defined as the (1D version of) the DSSCO. Appendix
A describes the wavefunction of this state for the simplest case,
$S=1/2$, as a Luttinger liquid with a Luttinger parameter that differs
from its value in other SRS ground states.

How can the DSSCO be extended to higher dimensions? In 2D, SRS can
remain broken at zero temperature ($T=0$), but is restored by thermal
fluctuations at any $T\neq0$ according to the Mermin-Wagner theorem.
Thus, the 2D DSSCO is a finite temperature phase and not a true quantum
ground state. A straightforward way to obtain it is to couple identical
chains ferromagnetically in the transverse directions, as shown in
Fig \ref{fig:DSSCO-2D}. Both the 1D and the 2D DSSCO are unstable
to infinitesimal random field disorder: $H_{dis}=\sum_{\boldsymbol{r}}\boldsymbol{h}(\boldsymbol{r})\cdot\boldsymbol{S}(\boldsymbol{r}),$
$\overline{\boldsymbol{h}(\boldsymbol{r})}=0$, $\overline{h_{a}(\boldsymbol{r})h_{b}(\boldsymbol{r}')}=h^{2}\delta_{ab}\delta(\boldsymbol{r}-\boldsymbol{r}')$,
$|h|\ll\mbox{all other coupling constants}$ and the overline denotes
a configuration average, because $\chi$ is an Ising order parameter
and $d=2$ is the lower critical dimension of the random field Ising
model \cite{Andelman1984}. In contrast, such disorder is a prerequisite
for the 3D generalization of the DSSCO. This is because, thermal fluctuations
cannot restore continuous symmetries in 3D, but quenched weak random
fields do, according to the Imry-Ma theorem \cite{Imry1975}. Analogous
ideas were discussed recently in the context of incommensurate charge
density waves (CDWs), that break continuous translational and discrete
rotational symmetries, in the pseudogap phase of the cuprates. The
analog of the DSSCO there was a \emph{vestigial nematic }phase, in
which chemical potential disorder acts as a random field for charge
density and restores translational symmetry while rotational symmetry
remains broken \cite{Nie2014}. The transition temperature for the
phase is finite, so the 3D DSSCO is a quantum ground state as well
as a $T\neq0$ phase. The $d$-dimensional version of the DSSCO is
thus naturally captured by the generalization of (\ref{eq:chi-1D}):
\begin{equation}
\chi_{d}=\frac{1}{L^{d}}\sum_{\boldsymbol{r}}\overline{\left\langle \boldsymbol{S}(\boldsymbol{r}-\hat{\mathbf{x}})\cdot\boldsymbol{S}(\boldsymbol{r})\times\boldsymbol{S}(\boldsymbol{r}+\hat{\mathbf{x}})\right\rangle }\label{eq:chi-d}
\end{equation}
$\chi_{d}$ obeys all the symmetries of the underlying lattice except
$x\to-x$ reflection. It is easy to check that translation and reflections
symmetries of the lattice prevent equilibrium current loops. Therefore,
the DSSCO lacks bulk orbital currents as well as spin moments and
consequently, lacks a density of total angular momentum expectation
values. The existence conditions of the DSSCO in various dimensions
are summarized in Table \ref{tab:Summary-of-conditions}, and its
symmetries in 3D are listed later in Table \ref{tab:Symmetry-table}.
Appendix B contains a simple toy model that is expected to realize
this phase as its ground state.

\begin{table}
\begin{centering}
\begin{tabular}{|c|c|c|}
\hline 
 & $T=0$  & $T\neq0$\tabularnewline
\hline 
\hline 
Clean  & 1D  & 2D\tabularnewline
\hline 
Dirty  & 3D  & 3D\tabularnewline
\hline 
\end{tabular}
\par\end{centering}

\caption{Dimensions in which the DSSCO can exist under various conditions.
$T$ is temperature and ``dirty'' refers to\emph{ }weak\emph{ }random
field disorder.\label{tab:Summary-of-conditions}}
\end{table}

\section{Experimental detection}

Most experiments that probe static TRS breaking, such as NMR and elastic
neutron scattering, explicitly measure local moments, so they cannot
see the DSSCO. What experiments \emph{can}?

As discussed earlier, spin moments are forbidden in the DSSCO by fundamental
properties of continuous symmetries, while mirror symmetries preclude
orbital moments involving other mobile degrees of freedom such as
itinerant electrons, if present. Unlike spin moments, though, orbital
moments only disobey discrete symmetries of the underlying lattice
(in addition to TRS). Thus, if sufficient mirror symmetries are broken,
for instance, by applying a suitable electric field or driving a current
through the system, current loops will generically form in the system
which can then be picked up by standard probes of magnetic order.
Explicitly, a straightforward symmetry analysis shows that the electromagnetic
response Lagrangian of the DSSCO contains a term $\mathcal{L}_{em}\sim\hat{\mathbf{Q}}\cdot\left(\boldsymbol{E}\times\boldsymbol{B}\right)$
upto coupling constants, where $\hat{\mathbf{Q}}\propto\chi$ is the
chiral ordering direction, so an external electric field induces a
magnetic response. Thus, a sharp signature of a DSSCO in 2D and 3D
will be the appearance of local moments in the presence of electric
fields or currents.

In 3D, another common experiment can sense the DSSCO without any other
fields for destroying mirror symmetries, namely, the polar Kerr effect:
the rotation of the plane of polarization of normally incident linearly
polarized light upon reflection. This effect requires vertical mirror
symmetries to be absent and, as long as linear response theory applies,
also needs broken TRS \cite{Halperin1992,Fried2014,Hosur2015E}. In
addition, either vertical reflections or time reversal combined with
horizontal translations must not be symmetry operations either. Usually,
these demands are met by bulk ferromagnetic moments perpendicular
to the reflection surface. In such systems, the sign of the effect
can be trained by a magnetic field and reverses upon flipping the
sample. In contrast, the DSSCO satisfies these conditions if the experiment
is performed on a low symmetry surface, such as the $(0kl),k\neq l$
surface of a cubic lattice with chiral ordering of the spins along
$x$. The effect originates from a net magnetic moment on the surface,
whose sign is determined by the bulk order parameter and the details
of the surface termination \footnote{Note that the appearance of surface moments does not contradict any
of the previous statements about a vanishing density of moments if
the sample is thermodynamically thick}. Thus, it cannot be trained by a magnetic field, and has the same
sign on opposite surfaces if the terminations are similar. Therefore,
it is strikingly different from the Kerr effect in most other systems.

In 1D, alternate ideas are needed to detect the DSSCO because current
loops are impossible and Kerr experiments are inapplicable. On the
other hand, the excitation spectrum contains gapped states corresponding
to domain walls of $\chi$, similar to the domain walls in a Ising
antiferromagnet, which are deconfined only in the 1D \cite{Pfeuty1970}.
A standard technique for probing magnetic domain walls is via inelastic
neutron scattering. Neutron spin couples linearly to electron spin,
so it creates a fluctuation in the magnetization (magnon) when it
scatters off a 1D Ising antiferromagnet. The magnon in turn decays
into a pair of domain walls, which leads to the neutron structure
factor exhibiting a characteristic ``2-particle continuum'' rather
than sharply defined magnon quasiparticles \cite{Oosawa2006,Coldea2010}.
Similarly, $\chi$ has the same symmetries as an ordinary current
and couples linearly to it, so electron diffraction off a 1D DSSCO
should show analogous signatures of domain walls in $\chi$. The details
of the experiment, though, are beyond the scope of this work.

\section{Application to the cuprates}

Recently, several families of the underdoped cuprates have been found
to exhibit a small polar Kerr effect in the pseudogap phase below
a temperature $T_{K}$ \cite{Xia2008,He2011,Karpetyan2012,Karpetyan2014}.
Assuming linear response, the effect indicates broken TRS below $T_{K}$
\cite{Halperin1992,Fried2014}. Unlike the effect in ferromagnets
and superconducting vortices, but like that in the DSSCO as discussed
earlier, its sign cannot be trained by a magnetic field and is the
same on opposite surfaces of the sample. These observations imply
that the effect does not stem from ordinary ferromagnetic moments
normal to the copper oxide planes. NMR experiments support this interpretation,
as Knight shift measurements below $T_{K}$ have set an upper bound
on the size of local magnetic moments that is two orders of magnitude
lower than that expected from some current proposals of TRS breaking
phases \cite{Varma1997,He2012}. To complicate matters further, the
sign of the effect also shows a ``memory effect'', i.e., it is unchanged
on heating to temperatures well above $T_{K}$ and cooling back, indicating
that some kind of order exists above $T_{K}$ but does not produce
a Kerr effect \cite{Xia2008}. Nernst effect data support this hypothesis,
as they see the $C_{4}$ symmetry of the copper-oxide plane broken
down to $C_{2}$ above $T_{K}$, but below the pseudogap temperature
$T^{*}$. Various X-ray scattering experiments have detected the onset
of incommensurate CDWs at $T_{K}$ \cite{Blackburn2013,Comin2015,LeTacon2014},
which suggests that the phase that forms above $T_{K}$ breaks only
some of the symmetries needed to produce a Kerr effect; the rest are
broken by the CDW. Finally, transmission experiments on thin films
indicate that the symmetries that are broken by the CDW are vertical
reflections \cite{Lubashevsky2014}. Below, a phenomenological (but
not microscopic) picture involving the DSSCO is presented in which
all the above experimental features can be accommodated and which
thus, may be relevant to the cuprates.

\begin{figure}
\begin{centering}
\subfloat[\label{fig:Phase-diagram}]{\begin{raggedright}
\includegraphics[clip,width=0.5\columnwidth]{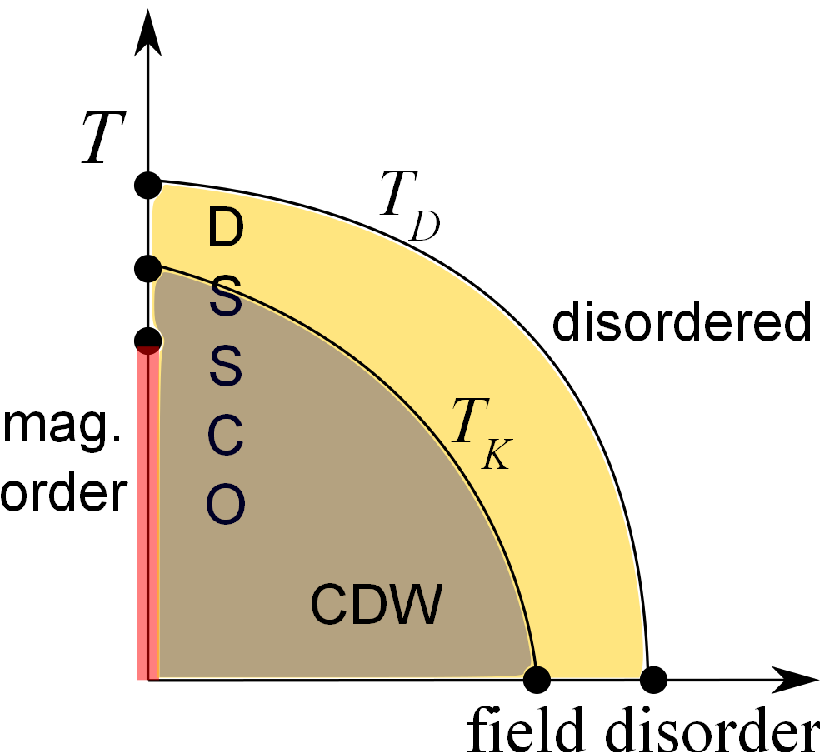}
\par\end{raggedright}

}\subfloat[\label{fig:Expt-setup}]{\begin{raggedright}

\par\end{raggedright}

\includegraphics[bb=100bp 50bp 420bp 300bp,clip,width=0.4\columnwidth]{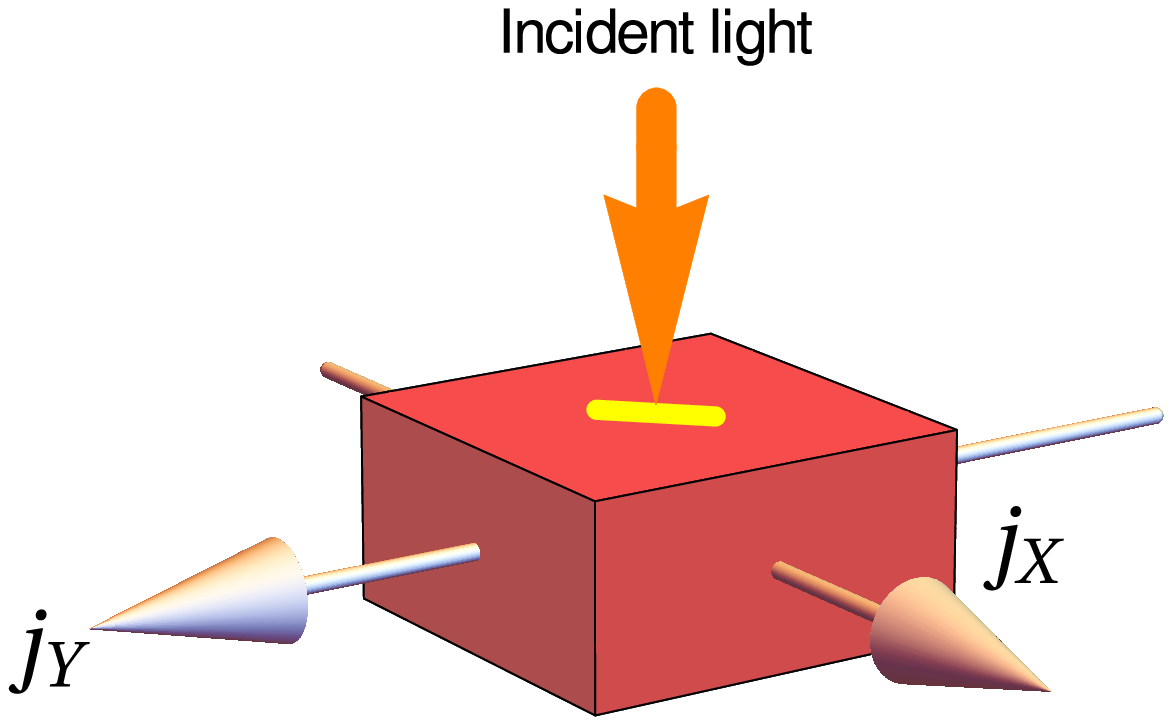}}
\par\end{centering}

\caption{(a) Schematic phase diagram that may be relevant to the cuprates.
The DSSCO forms at $T_{D}$ and coexists with charge order below $T_{K}$,
the Kerr onset temperature. $T_{D}$ may be $\approx T^{*}$; see
text for details. At zero disorder, it is not known \emph{a priori}
whether $T_{K}$ is higher or lower than the magnetic ordering temperature.
(b) Experimental setup for probing the DSSCO. For chiral ordering
along $x$, $j_{y}$ ($j_{x}$) would (would not) produce a Kerr effect
for reflection off the $xy$-surface.}
\end{figure}

Suppose the 3D DSSCO forms at a high temperature $T_{D}>T_{K}$ with
chiral ordering along $x$, one of the in-plane crystal axes. $C_{4}$
symmetry about the $z$-axis is then broken down to $C_{2}$, which
would give rise to an anisotropic Nernst effect. However, mirror symmetries
about the $xy$ and $xz$ planes, and the absence of static magnetic
moments, will suppress a Kerr effect and a Knight shift, respectively.
Next, suppose incommensurate CDWs that break all mirror symmetries
but respect twofold rotation symmetry about the $x$ or $y$ axis
onset at $T_{K}$. Such charge orders were discussed recently \cite{Hosur2013,Hosur2014}.
Below $T_{K}$, a Kerr signal is allowed by symmetry for reflection
off the $xy$ plane, and is likely to be small because it relies on
the formation of two orders -- the DSSCO and the CDW. Moreover, it
cannot be trained by a magnetic field and is invariant under flipping
the sample. This scenario involving two phase transitions can also
capture the memory effect. Specifically, the pattern of mirror symmetry
breaking by the CDW is likely determined by lattice defects. These
are extremely stable below the melting temperature of the solid, so
the sign of the Kerr effect will be the same as long as the $T<T_{D}$.
This scenario requires $T_{D}\gtrsim300K$ in underdoped YBa$_{2}$Cu$_{3}$O$_{6.5}$,
the temperature upto which the memory effect has been seen \cite{Xia2008}.
This is somewhat higher than the temperature below which the Nernst
effect saw $C_{4}$ symmetry breaking, $T^{*}\approx200$-$250K$
\cite{Daou2010}, but is still within some error bars, so it is not
unreasonable to suppose $T_{D}\approx T^{*}$. Below $T_{K}$, broken
TRS and mirror symmetry allow magnetic moments to form. However, these
moments will be small, possibly smaller than the NMR resolution, because
they depend on two orders. A simple test of the above picture would
be a Kerr signal between $T_{K}$ and $T_{D}$ triggered by a current
along one of the in-plane axes, but not along the other (see Fig.
\ref{fig:Expt-setup}). The signal, moreover, will flip on reversing
the current. Table \ref{tab:Symmetry-table} summarizes these symmetry
properties and Fig \ref{fig:Phase-diagram} shows a plausible phase
diagram.

\begin{table}
\begin{centering}
\begin{tabular}{|c|c|c|c|c|c|c|c|c|c|}
\hline 
 & TRS  & $M_{x}$  & $M_{y}$  & $M_{z}$  & $R_{x}^{2}$  & $R_{y}^{2}$  & $R_{z}^{2}$  & $\theta_{K}$  & KS\tabularnewline
\hline 
\hline 
DSSCO only & $\times$  & $\times$  & $\checkmark$  & $\checkmark$  & $\checkmark$  & $\times$  & $\times$  & $=0$  & $=0$\tabularnewline
\hline 
DSSCO+$j_{x}$ & $\times$ & $\times$ & $\checkmark$ & $\checkmark$ & $\checkmark$ & $\times$ & $\times$ & $=0$ & $=0$\tabularnewline
\hline 
DSSCO+$j_{y}$ & $\times$ & $\times$ & $\times$ & $\checkmark$ & $\times$ & $\times$ & $\times$ & $\neq0$ & $\neq0$\tabularnewline
\hline 
DSSCO+CDW  & $\times$  & $\times$  & $\times$  & $\times$  & $\checkmark$  & $\times$  & $\times$  & $\neq0$  & $\neq0$\tabularnewline
\hline 
\end{tabular}
\par\end{centering}

\caption{Symmetry properties of the DSSCO, chiral-ordered along $x$, in various
mirror-symmetry breaking fields. $M_{i}$ denotes $i\to-i$ reflection
and $R_{i}^{2}$ denotes $\pi$ rotation about the $i$ axis. The
CDW is assumed to respect (break) $R_{x}^{2}$ ($M_{x,y,z}$). $\theta_{K}$
is the Kerr angle for reflection off an $xy$-surface at normal incidence,
and KS denotes the NMR Knight shift.\label{tab:Symmetry-table}}
\end{table}

\section{conclusions}

In summary, the DSSCO is a novel phase of matter that violates TRS
but has no density of moments, unlike other TRS-breaking phases known
in condensed matter. It appears when a scalar chiral order of spins
partially melts, leaving behind residual broken TRS but unbroken continuous
SRS. A phenomenological picture, in which the DSSCO coexists with
a CDW, can be argued to have many of the features found in Kerr effect,
Knight shift, X-ray diffraction and Nernst effect experiments in the
pseudogap phase of the underdoped cuprates, and can be tested by looking
for a Kerr signal above $T_{K}$ on driving a suitable current through
the system. Whether the microscopics of this picture have any relevance
to the cuprates, however, is an open question. 
\begin{acknowledgments}
I thank Srinivas Raghu, Weejee Cho, Xiao-Liang Qi, Siddharth Parameswaran,
Yi Zhang, Brad Ramshaw and especially Steven Kivelson for insightful
discussions, Andrei Broido for helpful comments on the draft, and
the David and Lucile Packard Foundation for financial support.
\end{acknowledgments}

\bibliographystyle{apsrev4-1}
\bibliography{DSSCO_refs}

\appendix

\section{Wavefunction for $S=1/2$ chain \label{sec:Bosonization}}

For spin chains with SRS, the Lieb-Schultz-Mattis theorem implies
a gapless ground state if the spin per unit cell is a half-integer
\cite{Lieb1961}. It is generically a Luttinger liquid, so its effective
theory is one of free bosons: $\mathcal{H}_{Lutt}=\int\mathrm{d}x\left[K\left(\partial_{x}\phi\right)^{2}+\left(\partial_{x}\theta\right)^{2}/K\right]$.
Here, $K$ is the Luttinger parameter and $(\theta,\phi)$ are bosonic
fields satisfying $\left[\partial_{x}\phi(x),\theta(x')\right]=\left[\partial_{x}\theta(x),\phi(x')\right]=i\pi\delta(x-x')$,
and related to the spin variables as $S_{+}(x)\sim(-1)^{x}e^{i\theta(x)}$,
$S_{z}(x)\sim\partial_{x}\phi(x)$ in the simplest case of $S=\nicefrac{1}{2}$
\cite{SachdevQPT}. Under inversion ($x\to-x$) and time-reversal
($\boldsymbol{S}\to-\boldsymbol{S}$), $\theta$ is even upto constant
shifts while $\phi$ is odd, so $\mathcal{H}_{Lutt}$ does not couple
them. In such a theory, SRS fixes $K=2$.

In contrast, cross terms \emph{are }allowed in the DSSCO phase, and
the effective theory gets modified to 
\begin{equation}
\mathcal{H}_{Lutt}^{\chi}=\int\mathrm{d}x\left(\partial_{x}\phi,\partial_{x}\theta\right)\left(\begin{array}{cc}
K & -g\chi\\
-g\chi & 1/K
\end{array}\right)\left(\begin{array}{c}
\partial_{x}\phi\\
\partial_{x}\theta
\end{array}\right)
\end{equation}
to lowest order in $\chi$, where $g$ is a coupling constant. Explicitly,
the off-diagonal terms result from a mean field decoupling of a 6-spin
term in a suitable Hamiltonian: $-g\left[\boldsymbol{S}(x-1)\cdot\boldsymbol{S}(x)\times\boldsymbol{S}(x+1)\right]^{2}\to-2\chi g\boldsymbol{S}(x-1)\cdot\boldsymbol{S}(x)\times\boldsymbol{S}(x+1)$,
where $g>0$ favors the DSSCO. Equating $\left\langle S_{+}(x)S_{-}(0)\right\rangle $
and $\left\langle S_{z}(x)S_{z}(0)\right\rangle $ due to SRS yields
$K=2\sqrt{1-(g\chi)^{2}}<2$. In writing $\mathcal{H}_{Lutt}^{\chi}$,
terms $\propto e^{\pm4i\phi}$ have been dropped because they encourage
translational symmetry breaking, so are expected to be irrelevant
near the DSSCO fixed point. Their irrelevance is known for $K\leq2$
when $\chi=0$ \cite{SachdevQPT}, but a detailed renormalization
study is necessary to verify that it survives finite $\chi$.

\section{Toy Hamiltonians \label{sec:Toy-Hamiltonians}}

For integer spins per unit cell, the Lieb-Schultz-Mattis theorem places
no constraints on the ground state and it is generically gapped. The
gap guarantees short range entanglement \cite{Hastings2007} and therefore
amenability to description as a matrix product state. Equivalently,
it can be obtained in principle from an Affleck-Kennedy-Lieb-Tasaki
type Hamiltonian, which consists of a sum of projection operators
onto various spin channels acting on auxiliary spins \cite{Affleck1987}.
However, it is easier and more illuminating to write a classical Hamiltonian
that should yield the DSSCO for quantum spins, e.g., $H_{1D}=H_{bi}+H_{f\parallel}$,
where 
\begin{eqnarray}
H_{bi} & = & \sum_{i=1}^{2}K_{i}\sum_{x}\left[\boldsymbol{S}(x)\cdot\boldsymbol{S}(x+i)\right]^{2}\label{eq:Hbi}\\
H_{f\parallel} & = & -J_{\parallel}\sum_{x}\boldsymbol{S}(x)\cdot\boldsymbol{S}(x+3)\label{eq:Hfparallel}
\end{eqnarray}
represent biquadratic and ferromagnetic interactions along the chain,
respectively, with $K_{1,2},J_{\parallel}>0$. If the spins are classical
(large $S$), $H_{bi}$ mutually orthogonalizes every set of three
consecutive spins along $x$, and $H_{f\parallel}$ ensures that this
arrangement repeats along the chain, thus giving rise to the pattern
shown in Fig 1a of the main text. For small $S$, but $>1/2$, quantum
fluctuations partially melt the order and yield the DSSCO. If $S=\nicefrac{1}{2}$,
$\left[\boldsymbol{S}(\boldsymbol{r})\cdot\boldsymbol{S}(\boldsymbol{r}')\right]^{2}=\mbox{const.}-\boldsymbol{S}(\boldsymbol{r})\cdot\boldsymbol{S}(\boldsymbol{r}')/2$
and the biquadratic term reduces to exchange; hence, the above procedure
does not work and one is forced to start with a Hamiltonian with a
six-spin interaction $-g\left[\boldsymbol{S}(x-1)\cdot\boldsymbol{S}(x)\times\boldsymbol{S}(x+1)\right]^{2}$
to induce the DSSCO. For example, the modified $S=1/2$ Heisenberg
model
\begin{equation}
H_{1/2}=J\sum_{x}\boldsymbol{S}(x)\cdot\boldsymbol{S}(x+1)-g\left[\boldsymbol{S}(x-1)\cdot\boldsymbol{S}(x)\times\boldsymbol{S}(x+1)\right]^{2}
\end{equation}
is expected to have a DSSCO ground state for $g\gtrsim J$.

In $d$-dimensions, the discussion before Eq (2) of the main text
implies that the corresponding Hamiltonian for $S>1/2$ is $H_{dD}=\tilde{H}_{bi}+\tilde{H}_{f\parallel}+H_{f\perp}+H_{dis}$,
where 
\begin{eqnarray}
\tilde{H}_{bi} & = & \sum_{i=1}^{2}K_{i}\sum_{\boldsymbol{r}}\left[\boldsymbol{S}(\boldsymbol{r})\cdot\boldsymbol{S}(\boldsymbol{r}+i\hat{\mathbf{x}})\right]^{2}\label{eq:Hbi-1}\\
\tilde{H}_{f\parallel} & = & -J_{\parallel}\sum_{\boldsymbol{r}}\boldsymbol{S}(\boldsymbol{r})\cdot\boldsymbol{S}(\boldsymbol{r}+3\hat{\mathbf{x}})\label{eq:Hfparallel-1}\\
H_{f\perp} & = & -\sum_{a=y,z}J_{\perp a}\sum_{\boldsymbol{r}}\boldsymbol{S}(\boldsymbol{r})\cdot\boldsymbol{S}(\boldsymbol{r}+\hat{\mathbf{a}})\label{eq:Hfperp}\\
H_{dis} & = & \begin{cases}
0 & \mbox{in 1D and 2D}\\
\sum_{\boldsymbol{r}}\boldsymbol{h}(\boldsymbol{r})\cdot\boldsymbol{S}(\boldsymbol{r}) & \mbox{in 3D}
\end{cases}\label{eq:Hdis}
\end{eqnarray}
with $J_{\perp(y,z)}>0$ guaranteeing that the spin pattern is identical
on all $x$-directed chains. The appropriate Hamiltonian for $S=1/2$
is obtained simply by replacing $\tilde{H}_{bi}$ and $\tilde{H}_{f\parallel}$
by $H_{1/2}$, trivially generalized to $d$-dimensions:
\begin{equation}
\tilde{H}_{1/2}=J\sum_{\boldsymbol{r}}\boldsymbol{S}(\boldsymbol{r})\cdot\boldsymbol{S}(\boldsymbol{r}+\hat{\mathbf{x}})-g\left[\boldsymbol{S}(\boldsymbol{r}-\hat{\mathbf{x}})\cdot\boldsymbol{S}(\boldsymbol{r})\times\boldsymbol{S}(\boldsymbol{r}+\hat{\mathbf{x}})\right]^{2}
\end{equation}

\end{document}